\documentclass[sigconf]{acmart}

\setcopyright{acmlicensed}
\copyrightyear{2021}
\acmYear{2021}

\acmConference[EASE 2021]{Evaluation and Assessment in Software Engineering}{June 21--23, 2021}{Trondheim, Norway}
\acmBooktitle{Evaluation and Assessment in Software Engineering (EASE 2021), June 21--23, 2021, Trondheim, Norway}
\acmPrice{15.00}
\acmDOI{10.1145/3463274.3463349}
\acmISBN{978-1-4503-9053-8/21/06}

\newcommand{\logid}{\textsc{P\textsuperscript{3}}}

\begin{document}

\title{GDPR-Compliant Use of Blockchain for Secure Usage Logs}

\newcommand{\vzinst}{Technical University of Munich}

\author{Valentin Zieglmeier}
\affiliation{%
	\institution{\vzinst}
	\city{Munich}
	\country{Germany}}
\email{valentin.zieglmeier@tum.de}
\orcid{0000-0002-3770-0321}

\author{Gabriel Loyola Daiqui}
\affiliation{%
	\institution{\vzinst}
	\city{Munich}
	\country{Germany}}
\email{gabriel.loyola-daiqui@tum.de}
\orcid{0000-0002-9961-0866}

\begin{abstract}
The unique properties of blockchain enable central requirements of distributed secure logging: Immutability, integrity, and availability.
Especially when providing transparency about data usages, a blockchain-based secure log can be beneficial, as no trusted third party is required.
Yet, with data governed by privacy legislation such as the GDPR or CCPA, the core advantage of immutability becomes a liability.
After a rightful request, an individual's personal data need to be rectified or deleted, which is impossible in an immutable blockchain.
To solve this issue, we exploit a legal property of pseudonymized data:
They are only regarded personal data if they can be associated with an individual's identity.
We make use of this fact by presenting \logid{}, a pseudonym provisioning system for secure usage logs including a protocol for recording new usages.
For each new block, a one-time transaction pseudonym is generated.
The pseudonym generation algorithm guarantees unlinkability and enables proof of ownership.
These properties enable GDPR-compliant use of blockchain, as data subjects can exercise their legal rights with regards to their personal data.
The new-usage protocol ensures non-repudiation, and therefore accountability and liability.
Most importantly, our approach does not require a trusted third party and is independent of the utilized blockchain software.
\end{abstract}

\ccsdesc[500]{Security and privacy~Pseudonymity, anonymity and untraceability}
\ccsdesc[500]{Security and privacy~Privacy-preserving protocols}
\ccsdesc[100]{Computer systems organization~Peer-to-peer architectures}

\keywords{Blockchain, GDPR, Privacy, Anonymity, Logging}

\maketitle

\section{Introduction}

Blockchains are used for a multitude of applications. They can offer advantages in distributed contexts with untrusted participants, especially if immutability of data is required~\cite{zheng2017overview, lo2017evaluating}.
One potential application that benefits from these properties is distributed secure logging. Multiple secure logs based on blockchain were developed in recent years~\cite[e.g.,][]{ge2019permission, vanhoye2019logging, schaefer2019transparent}. The technology has many advantages for these applications, as it is a lightweight way to guarantee immutability of stored entries (integrity) and functions even in unreliable distributed networks (availability).

Yet, the unique properties of blockchain mean that the confidentiality of stored data cannot be guaranteed by default.
Even when storing minimal information, some form of identifier is required to denote ownership or association to entries. Without necessitating any information leak, the blockchain then allows any network participant to trace entries based on their identifier~\cite{reid2013analysis}.
At best, users can try to hide their association to blockchain entries by keeping their identifier secret and creating new addresses.
Even then, network participants can deduce information about users simply by analyzing publicly available data~\cite{androulaki2013evaluating}.
If the identifier is leaked or known to a third party, though, all respective entries can be retroactively associated with them.
This has been identified as a problem of blockchain-based secure logs~\cite{ge2019permission, schaefer2019transparent}, where confidentiality can be an important property~\cite{accorsi2010bbox}.
More critically, recent privacy legislation such as the General Data Protection Regulation (GDPR)~\cite{eu2016gdpr} of the European Union and the California Consumer Privacy Act (CCPA)~\cite{cali2018privacy} require those who store personal data to protect and, on request, even delete it.

One potential solution for this is to use permissioned (private) blockchains. Here, only parties with a legitimate interest in the stored data get access to the blockchain, managing it like any other database~\cite{ammous2016blockchain}.
This solves the issue of confidentiality (to a point) and can enable GDPR-compliance, but it removes the key advantage of a permissionless blockchain by requiring trust in those who can write to it~\cite{ammous2016blockchain}. Therefore, in this case there is no advantage of using blockchain compared to any other distributed database~\cite{lo2017evaluating}.

Intuitively, it seems as if this means a fundamental conflict between the requirements of privacy legislation and permissionless blockchain technology. Therefore, the only solution would be not to utilize blockchain when storing data that may be governed by the GDPR or CCPA.
We argue that this conflict can be solved differently, though. We want to combine the strengths of blockchain (such as immutability, decentralization, and fault tolerance~\cite{ge2019permission}) with the requirements of privacy legislation (confidentiality, deletability).

\paragraph{Use Case:} We are motivated by the idea of providing employees with more transparency regarding the usage of their data, e.g. by their employer.
Typically when personal data are handled, their usage is covered by privacy policies or company agreements. These policies are hard to read and understand~\cite{mcdonald2008cost}, calling into question whether individuals subjected to them truly understand their impact.
While some usages of their data might be beneficial to employees, giving these data up poses the risk of profiling and misusage.
To give individuals more oversight and control in situations of such asymmetric knowledge, various authors have described the idea to track individual usages of data and make the usage log available to data owners.
The concept has been discussed both in theory~\cite[see, e.g.,][]{brin1998transparent, agrawal2002hippocratic, weitzner2008information} and in practice~\cite[see, e.g.,][]{zyskind2015decentralizing, bagdasaryan2019ancile, schaefer2019transparent, zieglmeier2021trustworthy}.
This of course requires \emph{data owners} to trust that the usage log they are presented is complete and correct, necessitating a tamper-proof logging mechanism~\cite{schaefer2019transparent}.
Ideally, a trusted third party such as the employer is not required, as they might be interested to modify the log and potentially remove incriminating evidence.
To achieve this, \citeauthor{schaefer2019transparent} recently proposed utilizing a blockchain to serve as a secure usage log~\cite{schaefer2019transparent}.

\paragraph{Contribution:} Secure usage logs based on blockchain are fundamentally at odds with privacy legislation such as the GDPR. The metadata recorded in the blockchain are themselves personal data that need to be protected~\cite{demontjoye2014openpds}, an on request rectified or deleted~\cite{enisa2019recommendations}.
To enable secure and private logging of usage data with blockchain, we therefore contribute the pseudonym provisioning system \logid{}.
The system guarantees two properties: Proof of ownership, as required for deletion requests~\cite{enisa2021data}, and unlinkability, to provide users anonymity against adversaries~\cite{weber2012transaction}.
Most importantly, it does not require any changes in the utilized blockchain software.

\section{Pseudonymity and Anonymity}
\label{sec:pseudonymity-anonymity}

Two concepts are important to understand when discussing the applicability and implications of the GDPR: pseudonymity and anonymity.

Pseudonymized data are data where identifiers (such as names) have been replaced by pseudonyms, and the relationship between pseudonyms and identifiers is stored separately from the data themselves.
As the availability of this link allows re-identification, these data are not anonymous and fall under the provisions of privacy legislation~\cite{enisa2019recommendations}.
Anonymized data on the other hand are data that have been modified as such to make it impossible to re-identify an individual from them~\cite{iso25237pseudonymization, enisa2019recommendations}.
Importantly: While pseudonymous data are regarded personal data, anonymous data are not~\cite{enisa2019recommendations}. This means that to fulfill a user's legal right to erasure, we do not actually need to delete their personal data, as long as we can anonymize it by deleting their link to the pseudonym~\cite{enisa2019recommendations, enisa2021data}.

\section[P³: Private Pseudonym Provisioning]{\logid{}: Private Pseudonym Provisioning}

We present a pseudonym provisioning system to enable GDPR-compliant and secure data usage logging with blockchain.
As we have described above, the same datum can be considered anonymous and pseudonymous, depending on the knowledge of the respective party.
In the following, we will therefore deliberate the specific properties that the generated pseudonyms guarantee for each participating party.

First, we define our adversarial model and attacks that we consider. From our use case and the adversarial model, we derive requirements.
Then, we describe the \logid{} system and discuss how it addresses these requirements.

\subsection{Adversarial Model}
\label{sec:adversarial-model}

A user $u$ may be either in the set of \emph{data owners} $O = \{o_1,o_2,...,o_n\}$, in the set of \emph{data consumers} $C = \{c_1,c_2,...,c_m\}$, or both.
The adversary $\alpha$ can be any user and assume any role.
Whenever a consumer $c_i \in C$ accesses data of an owner $o_j \in O$, a usage $u_{ij}(c_i \to o_j)$ is appended to the usage log $U$ stored in the blockchain.

We assume that $\alpha$ has limited computational capacity, and can therefore never assume control over the blockchain network. Yet, within their means, they aim for maximum damage and therefore do not ``play fair''.
As we utilize blockchain, we know that attacks on the integrity of the log are infeasible for $\alpha$~\cite{zheng2017overview}.
Instead, $\alpha$ is motivated to attack the confidentiality of the stored data by gaining access to data that are not meant to be accessible by them.

Specifically, $\alpha$ tries to conduct the following attacks:
\begin{enumerate}
	\renewcommand{\theenumi}{\alph{enumi}}
	\item Derive from any entry $u_{ij}(c_i \to o_j)$ with $\alpha \notin \{c_i, o_j\}$ the identity of $c_i$ or $o_j$.
	\item Associate any two entries $u_{ij}(c_i \to o_j), u_{ik}(c_i \to o_k)$ with each other, thereby leaking their association with a single \emph{data consumer}.
	\item Associate any two entries $u_{ji}(c_j \to o_i), u_{ki}(c_k \to o_i)$ with each other, thereby leaking their association with a single \emph{data owner}.
	\item Leak the identity of $c_j$ for a stored usage $u_{jj}(c_j \to o_j)$ with $\alpha = o_j$, \emph{after} $c_j$ has legitimately exercised their right to erasure under the GDPR regarding $u_{jj}$.
	
\end{enumerate}

\subsection{Requirements}

\newcounter{Requirement}
\newcommand{\reqref}[1]{(\ref{#1})}
\newcommand{\stepreq}[1]{%
	\refstepcounter{Requirement}%
	\label{#1}%
}
\newcommand{\newreq}[1]{%
	\stepreq{#1}%
	\reqref{#1}%
}

From our use case and adversarial model arise five main requirements:
\newreq{req:tamper-proof} The usage log needs to be tamper-proof, preventing e.g. \emph{data consumers} from removing incriminating entries.
\newreq{req:no-trust}  No trusted third party is required.
\newreq{req:owners-verify} \emph{Data owners} can query for arbitrary log entries concerning their data and view their content. Importantly, they can prove the association of the \emph{data consumer} to the logged usage (non-repudiation). 
\newreq{req:consumers-verify} \emph{Data consumers} can query for arbitrary log entries concerning their usages and verify their content.
\newreq{req:confidentiality} No third party can derive their identity or usage information from data stored in the blockchain.

\stepreq{req:gdpr}
In addition, the data stored in the usage log are governed by the GDPR for as long as they can be associated with the identities of users. From that follows an additional requirement: After a legitimate request, (\ref{req:gdpr}a) a user's right to erasure~\cite[Art.~17]{eu2016gdpr} and (\ref{req:gdpr}b) a user's right to rectification~\cite[Art.~16]{eu2016gdpr} can be fulfilled.
We consider both issues a single requirement, as enabling the deletion of personal data indirectly enables rectification, by removing the incorrect entry and adding the rectified version.
Furthermore, as we have discussed in Section~\ref{sec:pseudonymity-anonymity}, we can fulfill a user's legal right to erasure by anonymizing their personal data. Therefore, we arrive at requirement \reqref{req:gdpr}: The ability to anonymize the personal data stored in the blockchain as such to make re-identification impossible.

\subsection[The P³ System]{The \logid{} System}

Our concept for the \logid{} system consists of four parts: The block structure, the new-usage protocol, the pseudonym generation algorithm, and the deployment architecture.
First, the block structure describes how usage data are stored in the blockchain and how they are protected.
Second, the new-usage protocol describes the interaction between \emph{data consumer} and \emph{data owner} when a datum is accessed and the usage is logged.
Third, the pseudonym generation algorithm enables the provisioning of private pseudonyms that guarantee unlinkability and proof of ownership.
Finally, the deployment determines the required trust and computational resources.

\subsubsection{Block Structure}

First, we define the actual data stored in each block of the blockchain.
As we have stressed above, our goal is to not require any changes to the underlying blockchain software, thereby allowing our solution to be used with existing blockchains.

As an example, we consider a new entry logging the usage $u_{ij}(c_i \to o_j)$.
Based on requirements \reqref{req:owners-verify} and \reqref{req:consumers-verify}, we need to store a one-time pseudonym for both the \emph{data consumer} and \emph{data owner}, to allow each party to query for entries concerning them. We will discuss the properties these pseudonyms guarantee in Section~\ref{sec:generation-algorithm}.
In addition, both parties need to be able to access the stored usage information, while preventing third parties from reading it, following \reqref{req:confidentiality}.

\begin{figure}[htbp]
	\centering
	\includegraphics[width=0.9\linewidth]{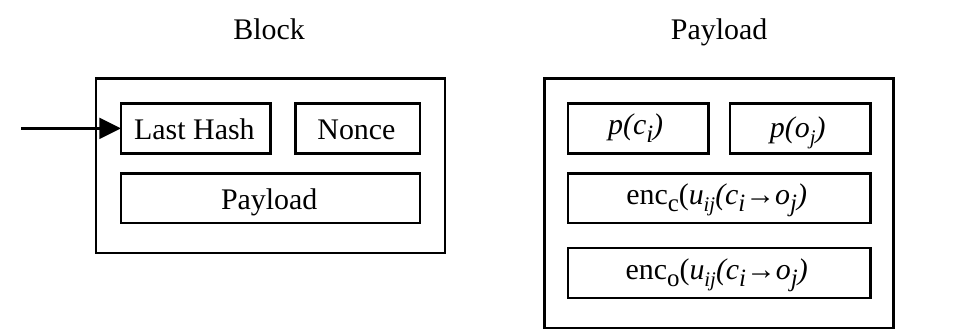}
	\caption{The components of each block (adapted from~\cite{nakamoto2008bitcoin}), for the exemplary usage $u_{ij}(c_i \to o_j)$. We replace the transaction history utilized in Bitcoin with a generic payload. The payload consists of the pseudonyms $p(c_i)$ and $p(o_j)$ of the \emph{data consumer} and \emph{data owner}, respectively. Furthermore, the logged usage is stored twice and encrypted once for the \emph{consumer} and once for the \emph{owner}.}
	\Description{Two boxes show the architecture of individual blocks in the blockchain as well as each payload. The block consists of the last hash, a nonce, and a payload. The payload consists of pseudonym(s) and encrypted data.}
	\label{fig:block}
\end{figure}

As shown in Figure~\ref{fig:block}, each block therefore contains a payload with the \emph{data consumer}'s pseudonym $p(c_i)$, the \emph{data owner}'s pseudonym $p(o_j)$, and two copies of the usage data.
Each copy is encrypted with an encryption function $enc()$, once with the \emph{owner}'s and once with the \emph{consumer}'s one-time public key. Importantly, this key is not shared with other parties and stored securely with the private key.
The chosen encryption should be regularly updated, but we require asymmetric (public-key) encryption~\cite{delfs2007public}.
As of this point, we recommend RSA~\cite{rivest1978method} with a key size of 3072 bits~\cite{lenstra2001selecting, kiviharju2017fog}.
Importantly though, each individual can choose their preferred key size, and therefore security level, themselves.

Besides allowing both parties to read it, storing the usage data twice is also important because the block is created by $o_j$ (see Section~\ref{sec:generation-protocol}). $c_i$ then needs to be able to verify the validity of the block. In case $o_j$ manipulates the stored entry, $c_i$ can utilize their copy of the usage and the non-repudiation evidence (see Section~\ref{sec:generation-protocol}) to defend themselves against the faked evidence.

Data usages often follow a similar or even the same form.
If a field is limited to few possible values, the deterministic nature of encryption algorithms will lead to the same ciphertext as long as the public key does not differ. This would allow a malicious user to group logs based on those fields, thereby reducing the confidentiality of the data.
In order to mitigate this potential threat, the used encryption method needs to support initialization vectors or optimal asymmetric encryption padding (OAEP).
An example for such an algorithm would be RSA OAEP~\cite{RSA_OAEP}.

\subsubsection{New-Usage Protocol}
\label{sec:generation-protocol}

Many properties we aim for hinge on the specific protocol that is followed when a data usage occurs.
Concretely, that means a \emph{data consumer} $c_i$ is accessing a datum $d_k$ of a \emph{data owner} $o_j$ and this usage being logged to the blockchain.
The protocol is designed to minimize the knowledge that each party has about the other party, while still ensuring the non-repudiability of their interaction.

Based on our use case, we consider the example of a department manager in the company accessing the completed tasks of one of their subordinates for a yearly report.
We specifically choose this example due to the inherent power asymmetry that exists between these parties.
Should the manager access data for improper purposes, they might be interested to use their power to remove any incriminating log entry.

The most important challenge is to guarantee non-repudiation of the occurred usage without a trusted third party. Therefore, we adapt the protocol designed by \citeauthor{markowitch1999probabilistic}~\cite{markowitch1999probabilistic} for our use case.
In short: After the protocol is successfully completed, each party will possess proof of their interaction with the other party. Importantly, $o_j$ can prove that $c_i$ has received the datum $d_k$.

\begin{figure}[htbp]
	\centering
	\includegraphics[width=0.85\linewidth]{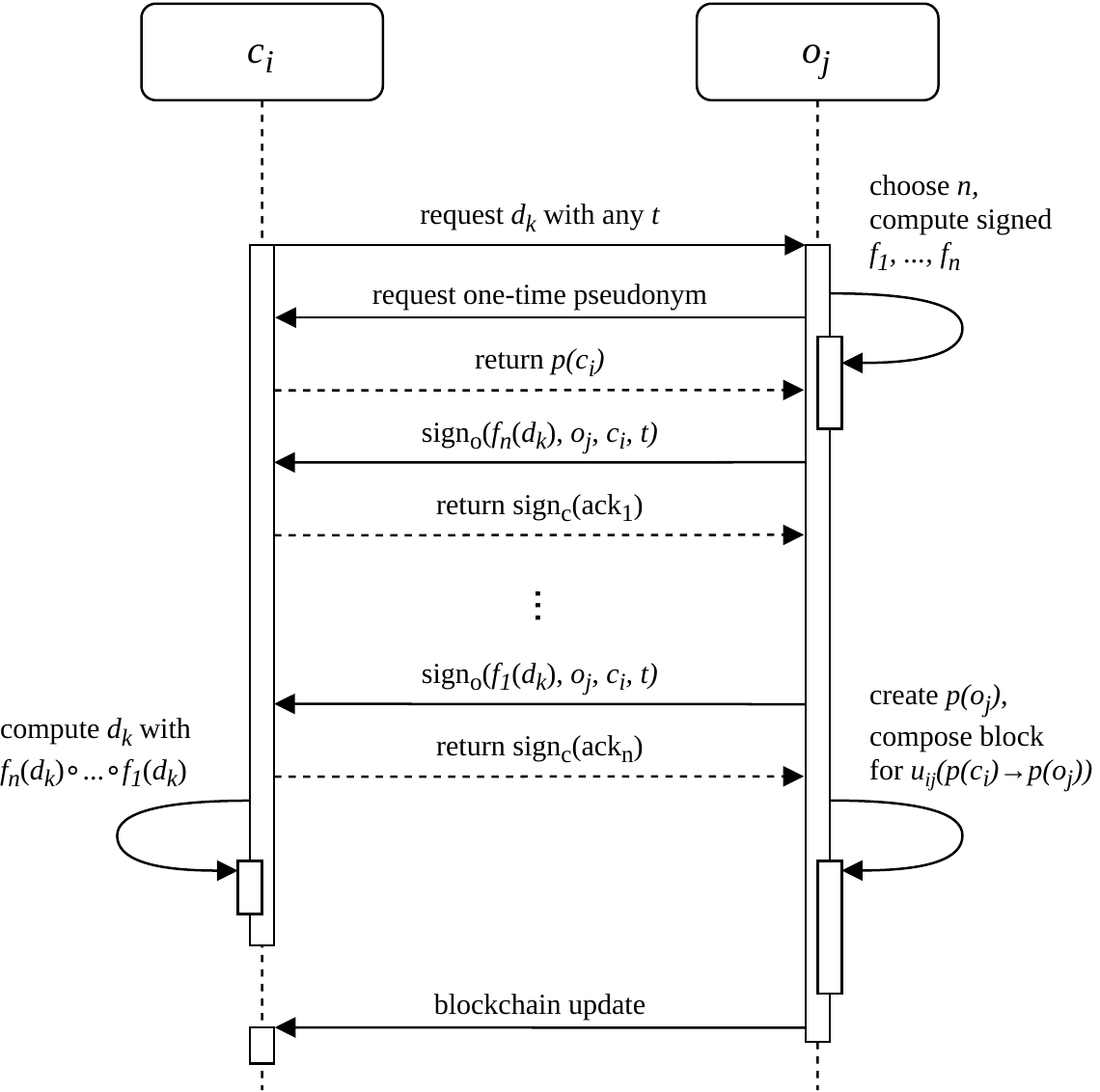}
	\caption{The new-usage protocol, adapted from~\cite{shamir1979share, markowitch1999probabilistic}. $c_i$ requests access to the datum $d_k$ from $u_j$ with an arbitrary value $t$ they choose. After receiving pseudonym $p_i$, the iterative non-repudiation protocol is performed, with a random number of steps $n$ unknown to $c_i$~\cite[see][pp.~5--6]{markowitch1999probabilistic}.
		Having received all messages, $c_i$ can compute the datum $d_k$ by composing the message payloads.
		Lastly, the block is created and distributed. Both parties store their respective non-repudiation evidence on their own machine.}
	\Description{A UML sequence diagram shows the steps of the protocol, as outlined in the caption.}
	\label{fig:protocol}
\end{figure}

The protocol is described in Figure~\ref{fig:protocol}. For this scenario, we assume that $o_j$ returns the requested datum $d_k$ directly.
Alternatively, they could also return the decryption key for a datum stored elsewhere.
The number of steps $n$ is chosen at random by $o_j$. For less critical data, it can be reduced~\cite[p.~7]{markowitch1999probabilistic}, e.g. to lower energy consumption or improve scalability.
Following \citeauthor{markowitch1999probabilistic}~\cite{markowitch1999probabilistic}, as $c_i$ cannot predict $n$, and if the chosen composition function takes long enough to compute, they will not be able to get any meaningful data when cheating~\cite[p.~5]{markowitch1999probabilistic}.
Only after having received all $n$ messages, they can compose them to receive the resulting datum~\cite{shamir1979share, markowitch1999probabilistic}.
Now, each party holds non-repudiation evidence of the interaction. For $o_j$, this is sign\textsubscript{c}(ack\textsubscript{n}), for $c_i$ it is the set of all messages sign\textsubscript{o}($f_x(d_k), o_j, c_i, t$) with $x \in (1,n)$~\cite[p.~5]{markowitch1999probabilistic}.

This protocol depends on the nodes being able to verify the authenticity of requests and, importantly, being protected against man-in-the-middle or eavesdropper attacks.
Therefore, each request is signed by the sender.
We do not aim to reinvent the wheel here, instead relying on the established HTTP over TLS standard~\cite{rfc2818httpovertls}.
This enables communication confidentiality and authenticity~\cite{krawczyk2013security}.
By utilizing the approach of a web of trust, as established in PGP~\cite{abdulrahman1997pgp}, nodes are fully independent of any trusted third party to verify certificates.
In that case, unknown certificates would be rejected and would need to be verified in-person.
Alternatively, if sensible for the specific deployment, a certificate authority can be used to sign the individual certificates used by each node to sign and encrypt its requests.
As these are often used in companies to enable the signing of internal emails or access to protected resources, no additional certification infrastructure is required in either case.

\subsubsection{Pseudonym Generation Algorithm}
\label{sec:generation-algorithm}

The central step towards our goal is the ability to generate unique one-time pseudonyms that guarantee unlinkability and enable proof of ownership, without requiring a trusted third party.
Unlinkability is required for the data we store to be able to qualify as anonymous data, as discussed in Section~\ref{sec:pseudonymity-anonymity}.
Proof of ownership, on the other hand, enables the owners of the pseudonyms to exercise their rights of erasure and rectification as given by the GDPR.

\citeauthor{florian2015sybil}~\cite{florian2015sybil} describe a pseudonym generation algorithm that serves as inspiration to our solution. Pseudonyms are guaranteed to be unlinkable to each other and to the real identity of the user. Furthermore, authenticity proofs enable proof of ownership, meaning that our requirements are met.
Beyond those properties, their algorithm provides sybil-resistance, which is achieved by requiring additional computational steps for joining a network and creating new pseudonyms~\cite[p.~68--69]{florian2015sybil}.
In our case, the additional property of sybil-resistance is not required, as there is no inherent danger in a user creating multiple pseudonyms (see Section~\ref{sec:generation-protocol}).
We can therefore omit these additional complexities and simplify our algorithm accordingly. By that, we reduce its computational complexity and energy consumption to a minimum.

Therefore, we define our pseudonym generation algorithm as follows:
As part of the new-usage protocol (see Section~\ref{sec:generation-protocol}), the user has created a new RSA private-public key pair with a key size of 3072 bits. As discussed above, the chosen encryption method can be updated if a higher level of security is appropriate.
Now, to generate the one-time pseudonym, the collision-resistant and cryptographic hash function BLAKE2~\cite{aumasson2013blake2}, specifically BLAKE2s~\cite[p.~121]{aumasson2013blake2}, is applied to create a cryptographic message digest.
Concretely, the user hashes the public key of their key pair, with the resulting irreversible and cryptographically safe digest representing their one-time pseudonym.
BLAKE2s ensures a digest size of at most 32 bytes, which is important to minimize storage requirements.

The pseudonym generated with our algorithm then guarantees three important properties:
First, the owner of the pseudonym, and only the owner, can prove the authenticity of the pseudonym (see below).
Second, the unique properties of the hash function guarantee collision-resistance and therefore uniqueness~\cite{applebaum2017low}.
Third, users only need to manage a single key pair for each block, significantly reducing the complexity of the operation and increasing its speed.

To enable proof of ownership, we make use of the asymmetric nature of the RSA key pair.
When a user wants to prove their ownership of a pseudonym, they sign a message with their private key and make available the corresponding public key.
The signed message proves that they are in possession of the private key.
The public key is then hashed by the recipient with the BLAKE2 hash function. If the result is the correct pseudonym, the ownership is proven.
As BLAKE2 is collision-resistant, it is infeasible for an attacker to guess a different string that would result in the same pseudonym.
Making matters even more secure, they would in addition need to crack the utilized RSA algorithm, as a collision-inducing string alone would not suffice to allow them to also sign a message with a matching private key.
We can therefore conclude that our algorithm is more than sufficiently secure.

\subsubsection{Deployment}
\label{sec:generation-deployment}

As we have hinted at above, the chosen deployment highly influences the privacy and security guarantees that can be given.
The protocol can be flexibly adapted and supports both centralized and peer-to-peer architectures.
We aim for our solution to not be dependent on any trusted third party. Therefore, our deployment architecture is fully decentralized.

\begin{figure}[htbp]
	\centering
	\includegraphics[width=0.9\linewidth]{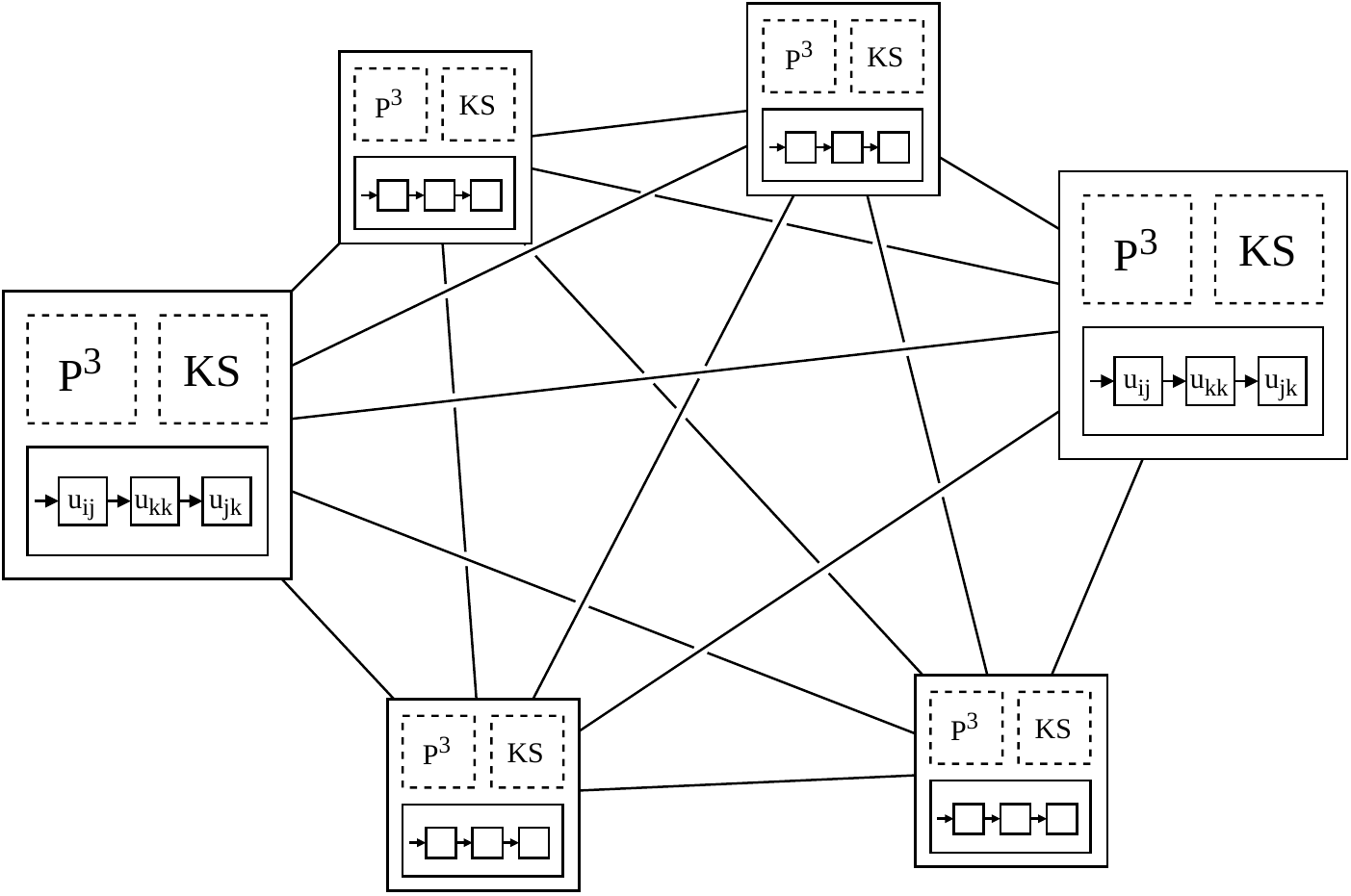}
	\caption{Deployment in a fully decentralized peer-to-peer architecture. Each node runs its own \logid{} instance and private key store KS, while the blockchain copy is shared within the network~\cite[see also][]{nakamoto2008bitcoin}.}
	\Description{Six boxes representing individual nodes are shown, all interconnected. Each box contains three components, the \logid{} component, the KS component, and the blockchain.}
	\label{fig:architecture}
\end{figure}

The central component of the deployment is the \logid{} module that handles pseudonym provisioning and key management.
Each node in the peer-to-peer network runs its own \logid{} instance as well as a private key store KS to store its RSA key pairs and pseudonyms (see Sections~\ref{sec:generation-protocol} and \ref{sec:generation-algorithm}).
The usage log blockchain is permissionless and therefore shared between all nodes participating in the network.
Therefore, the architecture is fully decentralized, with each node communicating directly with other nodes (see Figure~\ref{fig:architecture}).

The management of a large number of keys as required by our approach can itself become a privacy risk. Therefore, the chosen key management strategy of the key store component KS is relevant. Applying the Master \& Sub Key Generation pattern~\cite{liu2020design} reduces key management complexity while maximizing privacy~\cite[p.~5]{liu2020design}.
To additionally protect the master key, the Key Shards pattern~\cite{liu2020design} can be utilized by users to protect against key loss~\cite[p.~7]{liu2020design}.

In theory, the nodes would not necessarily need to store their key pairs and used pseudonyms at all to reduce their attack surface.
Yet, these are important to enable requirements \reqref{req:owners-verify}, \reqref{req:consumers-verify}, and \reqref{req:gdpr}.
To be able to query for entries concerning their usages, users need to know which pseudonyms belong to them. In theory, they could iterate all blocks in the blockchain and simply try to decrypt their content, but this quickly becomes infeasible.
Additionally, to exercise their GDPR-awarded rights, users need to prove their ownership of a pseudonym, requiring them to be in possession of the respective private key for the specific block (see Section~\ref{sec:generation-algorithm}).
Still, each user can choose on their own how to manage their keys. If they prefer the maximum level of security, they are free to delete new keys immediately after usage.

\section{Analysis}

In the following, we analyze our concept based on three core aspects: Its compliance with the GDPR (our central requirement), its robustness against attacks, and its protocol security.

\subsection{GDPR Compliance}

We have discussed above (see Section~\ref{sec:pseudonymity-anonymity}) that data can be considered pseudonymous and anonymous. When a possibility for re-identification exists, they count as pseudonymous and therefore fall under the provisions of the GDPR.
To comply with the GDPR, we need to enable data subjects to exercise their right to erasure as well as their right to rectification.
We have also deliberated that by enabling the right to erasure, the right to rectification is also given, as we can simply delete the incorrect block and add a new, corrected version.

Each block in our approach gets its own transaction pseudonym. We have shown in Section~\ref{sec:generation-algorithm} that these pseudonyms are unlinkable to the individual's identity and to each other.
For a recorded usage $u_{ij}(p_x(c_i) \to p_y(o_j))$, only $c_i$ and $o_j$ know the association of the other party's single one-time pseudonym ($p_x$ or $p_y$) to their real-world identity.
In fact, $c_i$ and $o_j$ have to prove their identity to each other in the first step of the protocol (see Section~\ref{sec:generation-protocol}).
Due to the guaranteed unlinkability, this link is only given for the single pseudonym created for that block, i.e. $o_j$ only knows the link $p_x(c_i) \leftrightarrow c_i$.
This case is covered by the existing GDPR provisions.
The association is not stored in the blockchain, but only on the nodes of $c_i$ and $o_j$. That means the request for deletion simply has to be forwarded to them.
Considering an adversarial user, they might just not fulfill that request for deletion of course (see also Section~\ref{sec:analysis-attacks}).
At first glance, this could imply that our protocol does not offer an advantage over just modifying the blockchain and asking all nodes to delete their old copy.
Importantly, though, we do not deal with \emph{unknown} nodes.
When a deletion request is raised, the data subject \emph{knows} the identity of the offending user, and can \emph{prove} it (see Section~\ref{sec:generation-protocol}).
That means, the individual can then be made responsible for deletion under the GDPR, and can be sued in case they do not follow through.

As soon as the association of the individual's identity to the pseudonym has been deleted, the data stored in the blockchain are anonymized.
Then, they do not qualify as personal data anymore (see also Section~\ref{sec:pseudonymity-anonymity}), satisfying the right to erasure.

\subsection{Robustness Against Attacks}
\label{sec:analysis-attacks}

We have described an adversarial model in Section~\ref{sec:adversarial-model}, with adversary $\alpha$ trying to subvert the confidentiality of the stored logs.
Specifically, they try to conduct four attacks. For each, we will discuss the robustness of our approach against the attack and potential implications.

Firstly, $\alpha$ tries to (a) derive from any usage $u_{ij}(c_i \to o_j)$ the identity of $c_i$ or $o_j$.
As the transaction pseudonyms for both $c_i$ and $o_j$ are created with the same algorithm, this attack depends on being able to reverse the employed pseudonym generation. The one-time cryptographic security of our utilized algorithm has been shown~\cite[see, e.g.,][]{aumasson2013blake2, luykx2016security}, guaranteeing it to be irreversible.
Furthermore, as each transaction gets its own pseudonym, each would have to be cracked independently, meaning an exponential effort.

Secondly, $\alpha$ tries to (b) associate any two usages $u_{ij}(c_i \to o_j)$, $u_{ik}(c_i \to o_k)$ with each other, revealing their association with a single \emph{consumer} and (c) associate any two usages $u_{ji}(c_j \to o_i)$, $u_{ki}(c_k \to o_i)$ with each other, thereby leaking their association with a single \emph{owner}.
We can discuss both attacks together, as they hinge on the same security mechanism.
The cryptographic security of our algorithm is guaranteed by the cryptographic security of the two underlying algorithms, which has been shown for both~\cite[see][]{mahto2016security, luykx2016security}.
Therefore, this again depends on the ability of $\alpha$ to reverse the transaction pseudonym generation. As we have shown above, this can be considered infeasible.

Finally, $\alpha$ tries to (d) leak the identity of $c_j$ for a stored usage $u_{jj}(p(c_j) \to p(o_j))$ with $\alpha = o_j$, after $c_j$ has exercised their right to erasure.
In the last section, we have detailed that the association of $c_j$ to their pseudonym is known to $o_j$ for blocks storing usages of data that $o_j$ owns.
We have no way to technically force $o_j$ to delete this association when $c_j$ exercises their right to erasure.
Now, assuming $o_j$ does not delete this data and wants to utilize their knowledge, e.g., by publishing the real identity of $c_j$ and their one-time pseudonym $p(c_j)$. By itself, this proves nothing though, as there is no technical relationship between the pseudonym and the identity of $c_j$ (see Section~\ref{sec:generation-algorithm}).
To actually prove the association of $c_j$ to $u_{jj}$, $o_j$ therefore needs to publish their non-repudiation evidence (see Section~\ref{sec:generation-protocol}).
As shown by \cite{markowitch1999probabilistic}, this evidence by design contains their own identity (through their signature) as well~\cite[pp.~5--6]{markowitch1999probabilistic}.
This means that $o_j$ would automatically also leak their own identity, making them legally liable.

As this scenario is covered by the privacy legislation and can be prosecuted accordingly, we consider it a non-issue for most cases.
Still, for the most secretive of environments, this might not be enough of a guarantee.

\subsection{Protocol Security}

For the highest-security deployments, our peer-to-peer architecture enables pseudonym generation and block creation without necessitating a trusted third party, mitigating most attack vectors on the integrity and confidentiality of data.
Two potential attack vectors on the confidentiality of the exchanged information remain: The messages sent on block creation, and the block update.

Firstly, when creating a new block for a usage $u_{ij}(c_i \to o_j)$ following the protocol (see Section~\ref{sec:generation-protocol}), communication between $c_i$ and $o_j$ has to occur. Even though HTTP over TLS is utilized, which prevents $\alpha$ from listening in as an eavesdropper~\cite{krawczyk2013security}, they may still deduce that there is some usage association between the nodes.
There are various options to address this. In our view, the easiest to implement would be to add an additional fake protocol. This would work much the same way, only that a special non-existent datum $d_0$ is requested. Then, both parties understand that this is just a fake request, and no actual block is added to the blockchain.
This fake protocol can be run by nodes in randomized intervals, choosing arbitrary other nodes to request $d_0$ from. Thereby, we can hide real requests in the noise of these fake requests.

What remains then is the block creation. Even if fake protocols are run regularly, $\alpha$ could simply watch for blockchain updates and derive from those which two nodes were responsible for the new block.
The simplest mitigation of this is to add a random wait before the block is added.
Then, plausible deniability is enabled, as there are a sufficient number of other potential users that might have been responsible. In fact, nodes might even wait for a certain number of block updates before publishing their update. Here, too, each node can decide itself the level of confidentiality it requires, and act accordingly.
Furthermore, traditional blockchain algorithms already (indirectly) protect from $\alpha$ understanding the originator of a blockchain update. As the architecture is designed to be peer-to-peer, the mere fact that a node sends a block update does not give $\alpha$ any additional information about its creation. As nodes forward block updates to other nodes, the specific node that sends $\alpha$ the update might simply have forwarded it~\cite{nakamoto2008bitcoin, zheng2017overview}.

This shows that our protocol can be adapted to fulfill the highest requirements towards information security.

\section{Related Work}

Various proposals for how to solve the conflict between the GDPR requirements and blockchain properties exist.
In the following, we describe important works and discuss how they differ from our approach. The overview by~\citeauthor{pagallo2018chronicle}~\cite{pagallo2018chronicle} serves as a foundation. A recent systematic literature review~\cite{haque2021gdpr} confirms its completeness.

\subsection{Key Destruction}

A trivial solution is to encrypt all data that are stored on the blockchain and delete the decryption key if the data are to be deleted, as discussed by~\citeauthor{pagallo2018chronicle}~\cite{pagallo2018chronicle}.

While it is easy to implement, this approach is flawed.
Encryption per se only guarantees pseudonymity of data~\cite{eu2016gdpr, enisa2021data}, therefore the data protection requirements still apply~\cite{enisa2021data}.
More problematically, though, if the full content of the block is encrypted, querying history becomes all but impossible, which is a requirement in secure logs for efficiently reading past entries. Only the affected parties themselves would be able to retrieve their entries, and only with high computational overhead by going through every block and trying to decrypt it.

Our proposal in contrast enables querying of entries based on the one-time pseudonyms, while enabling retroactive anonymization of data, which fulfills the requirements of the GDPR's right to erasure~\cite{enisa2019recommendations, giessen2019blockchain}.

\subsection{Forgetting Blockchain}	

\citeauthor{farshid2019design} propose to achieve a GDPR-compliant blockchain by automatically deleting blocks from the blockchain after a certain amount of time has passed~\cite{farshid2019design}.

As the described network no longer contains a genesis block, joining a network becomes a challenge. The authors propose to ask other nodes for the current block and just accept it if all the returned blocks are equal~\cite{farshid2019design}. Since there is no way to verify that the received block reflects the true state of the network, joining it requires trust and does not satisfy the integrity constraint.
Secondly, the nature of their approach prevents the existence of a chain history. Applications relying on the full history, specifically in the case of secure logs, would therefore not work with this algorithm.
Furthermore, this proposal only achieves eventual GDPR compliance, since a block can only be deleted after the predefined time has passed. If a user requests deletion of their data, this request cannot be fulfilled immediately.
For this reason, it is questionable if the presented idea is compatible with the GDPR.
Most problematically though, the data are only actually deleted if all nodes behave honestly~\cite{farshid2019design}. Any node can simply decide not to delete older blocks, meaning that no additional privacy guarantees can be given.

In contrast to the forgetting blockchain, our proposal does not require adaptation of the utilized blockchain software and is therefore easier to integrate into existing blockchains. Furthermore, we do not depend on the honesty of \emph{arbitrary} and \emph{unknown} nodes. In contrast, only one \emph{known} party has \emph{provable} access to additional identity information and can be held liable under the GDPR.

\subsection{Redactable Blockchain}

The reason that the immutability of data stored in a blockchain can be guaranteed is the utilized hash function: An ideal hashing algorithm guarantees hashes that are one-way, which means impossible to reverse, and collision-free.
Then, blocks cannot be replaced without notice, as any change would result in a new hash, thereby invalidating the chain.

Redactable blockchains utilize so-called \emph{chameleon} hash functions to determine the hash of a block. Such hash functions are collision-resistant as long as a secret known as \emph{trapdoor} is not known. If one is in possession of said secret, they can efficiently compute colliding hashes~\cite{redactableBlockchain, trapdoorHash}.
With the power to create hash collisions, any block can be replaced or even removed~\cite{redactableBlockchain}, making the blockchain effectively arbitrarily editable.

In order to function, such a redactable blockchain network needs a trusted third party that is in possession of the trapdoor and can decide which block to edit~\cite{redactableBlockchain}. This constraint again requires trust, thereby calling into question the value of utilizing blockchain at all~\cite{pagallo2018chronicle}.
Furthermore, similar to the forgetting blockchain, every individual node needs to be trusted. Redactions are published as chain updates, allowing \emph{arbitrary} nodes to make a copy of the removed or edited entry before updating their chain~\cite{pagallo2018chronicle, giessen2019blockchain}.
This means that, effectively, no privacy guarantees can be given.

Our solution on the contrary does not require a trusted third party and functions even in the face of adversarial nodes.

\subsection{Hashing Out}

Hashing out refers to the practice of saving the hash of the data on the blockchain and the data itself off-chain~\cite{pagallo2018chronicle}.
This approach is one of the most commonly used ideas to guarantee GDPR compliance in blockchain solutions~\cite[see, e.g., ][]{privacyByBlockchainDesign, schaefer2019transparent, vanhoye2019logging}. This is because the on-chain hash does not contain any private or personal data and the off-chain data can be deleted or modified to comply with a data subject's request.

There are two major downsides, though. Since the data are not stored on the blockchain, this solution is not truly decentralized and requires trust in the authority managing the data~\cite{ibanez2018blockchains}.
Furthermore, using this approach one can only be sure of the existence of entries, not of their content.
That means it is effective only as long as the log is not tampered with, but does not guarantee accountability or non-repudiation.

We allow users to benefit from the maximum possible utility of the log data, specifically giving them access to the content of stored log entries that relate to them, while still guaranteeing the same level of confidentiality. Most importantly, we require no trusted third party.

\section{Discussion}

Even with the best technical protections, individual users remain as an often-abused attack surface~\cite{wl_InsiderThreat, wl_Deception, wl_Revisited, wl_Psych}.
For most data that we store, there is no danger of users unwillingly leaking information about other parties except for themselves, with one exception: \emph{data owners} could be tricked or hacked to reveal the identities of \emph{consumers} of their data.
In our current architecture, it is impossible to prevent this case, yet we consider the attack surface to be acceptably small. To get access to a meaningful dataset about the usage pattern of a \emph{data consumer}, an adversary would have to hack or phish each individual \emph{data owner} of data accessed by said \emph{data consumer}. We consider that infeasible.

Furthermore, we prioritize security. That in turn means that other properties, such as availability or scalability, are not optimized for.
Regarding availability, each individual node manages its own data and has to be reached when accessing data. Should the node crash, be shut off, or otherwise disconnected from the network, the \emph{data consumer} is prevented from continuing their work.
In scenarios where the availability of the nodes is prioritized higher than their security and independence, we can imagine running user's nodes, e.g., on virtual servers. While this adds an attack surface and removes control from the user, it can improve availability.
Considering scalability on the other hand, the requirement to have a key pair for each individual access can quickly lead to a huge key store. This is one of the biggest weaknesses of our approach. Beyond pooling requests, the most sensible solution in less critical scenarios might be to reduce key size or reuse pseudonyms.
As always, this is a trade-off depending on the specific requirements.

To expand on this point, there has been broader discussion on what constitutes ``good enough'' software security and how to make objective judgments about it~\cite{tondel2020achieving}.
\citeauthor{tondel2020achieving} suggest to not only consider the system from the perspective of the adversary (as we have done), but to additionally factor in other perspectives such as those of users or operators~\cite[p.~364]{tondel2020achieving}.
Following their proposal, it might therefore be sensible to conduct a broader analysis of the system that also covers these perspectives before deploying it.
This could be important to ensure user acceptance and usability, as well as to address potential practical issues with deployment and operation, which might otherwise hinder adoption.

\section{Conclusion}

At first glance, the requirements of privacy legislation such as the GDPR seem to fundamentally contradict the technical properties of blockchain.
This makes it seemingly impossible to store data governed by the GDPR in a blockchain.
Yet, the unique advantages of utilizing blockchain technology can be beneficial in cases where a tamper-proof secure log is needed, such as when logging data usages. By design, such a secure usage log stores personal data governed by the GDPR.
To solve this conflict, we have analyzed the requirements of the GDPR and developed a concept to address them.
Our concept exploits a specific legal property of pseudonymized data: Depending on if they can be linked to the identity of a user, they count as personal data or as anonymous.
We describe a pseudonym generation algorithm and protocol that guarantee unlinkability and anonymity, while enabling proof of ownership and authenticity.
With those properties, the requirements of the GDPR can be fulfilled.
Our block structure and deployment architecture then allow individuals to query for and read arbitrary usage log entries concerning their data, while protecting them from attacks on their confidentiality by adversaries.
In our analysis, we deliberate how this allows us to fulfill the requirements of the GDPR by enabling confidentiality and the rights to erasure and rectification, while at the same time benefiting from the properties of blockchain, specifically guaranteeing the integrity of the logged data.
Related works require either the use of a permissioned blockchain, necessitating a trusted third party, or modifying the utilized hashing algorithms or blockchain software to make the blockchain mutable. Both approaches entail effectively giving up the advantages of blockchain, thereby calling into question the use of blockchain in the first place.
With the \logid{} system, GDPR-compliant secure usage logs based on blockchain can be implemented. In addition, our approach does not require a trusted third party, can be used with existing blockchain software without requiring changes, and is secure and flexible enough to enable deployment even in highest-security settings.

\begin{acks}
This work was supported by the German Federal Ministry of Education and Research (BMBF) under grant no. 5091121.
\end{acks}

\bibliographystyle{ACM-Reference-Format}
\bibliography{references}

\end{document}